\documentclass[aps,prb,twocolumn,groupedaddress,showpacs,showkeys,floatfix]{revtex4}
\usepackage{graphicx}
\pdfoutput=1

\begin{document}
\preprint{}


\title{Dynamic fracture of icosahedral model quasicrystals:\\
A molecular dynamics study}
\author{Frohmut R\"osch, Christoph Rudhart, Johannes Roth, and Hans-Rainer
  Trebin}
\affiliation{Institut f\"ur Theoretische und Angewandte Physik, Universit\"at
  Stuttgart, Pfaffenwaldring 57, 70550 Stuttgart, Germany}
\author{Peter Gumbsch}
\affiliation{Institut f\"ur Zuverl\"assigkeit von Bauteilen und Systemen,
Universit\"at Karlsruhe, Kaiserstr.~12, 76131 Karlsruhe, Germany}
\affiliation{Fraunhofer Institut f\"ur Werkstoffmechanik,
 W\"ohlerstr.~11, 79108 Freiburg, Germany}



\begin{abstract}
  Ebert et al.\ [Phys.\ Rev.\  Lett.\ 77, 3827 (1996)] have fractured
  icosahedral Al-Mn-Pd single crystals in ultrahigh vacuum and have investigated the
  cleavage planes in-situ by scanning tunneling microscopy (STM). Globular patterns in the
  STM-images were interpreted as clusters of atoms. These are significant
  {\em structural} units of quasicrystals. The experiments of Ebert et al.\ 
  imply that they are also stable {\em physical} entities, a property
  controversially discussed currently.  For a clarification we performed the
  first large scale fracture simulations on three-dimensional complex binary
  systems.  We studied the propagation of mode I cracks in an icosahedral
  model quasicrystal by molecular dynamics techniques at low temperature. In
  particular we examined how the shape of the cleavage plane is influenced by
  the clusters inherent in the model and how it depends on the plane structure.
  Brittle fracture with no indication of dislocation activity is observed.
  The crack surfaces are rough on the scale of the clusters, but exhibit
  constant average heights for orientations perpendicular to high symmetry
  axes.  From detailed analyses of the fractured samples we conclude that
  both, the plane structure and the clusters, strongly influence dynamic
  fracture in quasicrystals and that the clusters therefore have to be
  regarded as physical entities.
\end{abstract}
\maketitle

\narrowtext


\section{INTRODUCTION}
\label{introduction}

Quasicrystals are intermetallic compounds with long-range quasi-periodic
translational order. They possess well-defined atomic planes and hence
diffract electromagnetic and matter waves into sharp Bragg spots. But they
also display atomic clusters as basic building blocks \cite{Henley91,Elser96},
whose arrangement in space is compatible with the planar structure. These
clusters consist for example of several shells of icosahedral symmetry
(\mbox{Bergman-}, \mbox{Mackay-}, pseudo-Mackay-clusters). Or they form
polytopes, e.g.\ decagonal prisms, which like the unit cells of periodic
crystals fill space, although with large overlaps (``quasi-unit-cell
picture'') \cite{Steinhardt98}. Janot and others
\cite{Janot94,Janot97,Jeong94} have postulated that a self-similar
hierarchical assembly of the clusters is responsible for the stability of
quasicrystals and for many physical properties, like the low electric
conductivity. However, it is a controversial and persistent discussion,
whether the clusters are merely structural units or whether they represent
physical entities. The discussion was fueled by an experiment of Ebert et
al.~\cite{Ebert96}, where icosahedral Al-Mn-Pd was fractured under ultrahigh
vacuum conditions at room temperature. Scanning tunneling microscopy images of the
cleavage planes revealed elements of 0.6 to 1 nm in diameter
\cite{Ebert03}. The authors argue that these are the clusters which were
circumvented by the crack and hence form highly stable aggregates of matter.
Others point out that flat terraces evolve on fivefold surfaces of i-Al-Mn-Pd 
when annealed at high
temperatures~\cite{Schaub94,Gierer97,Cappello02,McGrath02}. As this requires
truncated rows of Bergman and Mackay clusters it is stated~\cite{Kasner03}
that these therefore could not represent firm entities.

In the present article we report on molecular dynamics simulations of crack
propagation in a three-dimensional icosahedral model quasicrystal at low
temperature. Seed cracks are inserted along different planes and therein
along different directions. The fracture planes are carefully analysed to
answer the role of clusters in dynamic fracture.
In Sec.~\ref{fracture} we provide some requirements on the theoretical
description of fracture. In Sec.~\ref{modelnmethod}
the model quasicrystal, the molecular dynamics technique, and
the methods to visualise the results of the simulations are outlined.
Subsequently, in Sec.~\ref{results} the simulation results are presented and
then discussed in Sec.~\ref{discussion}.



\section{FRACTURE}
\label{fracture}

The stress concentration and the strength of the loading at a crack
tip are determined by the macroscopic geometry and dimensions of a sample.
In linear elastic continuum mechanics, a sharp mode I (opening
mode) crack is characterised by a singular stress field and a corresponding
displacement field, which both are proportional to the stress intensity factor
$K$. This factor is proportional to the applied external load and contains the
geometry of the sample. A simple energy based condition for crack propagation
is the Griffith criterion~\cite{Griffith21}. It states that a crack is in
equilibrium when the
change in mechanical energy per unit area of crack advance -- the energy
release rate $G$ -- equals the change in surface energy of the two fracture
surfaces, $2 \gamma$. In continuum mechanics the energy release rate is
proportional to the square of the stress intensity factor for a given mode. A
crack then should start moving when the stress intensity factor exceeds the
critical Griffith value.

A continuum mechanical description of fracture, however, has a few
drawbacks. First, the requirements for linear elasticity are no longer valid
near the crack tip where atomic bonds clearly become non-linear and eventually
break. Second, a continuum theory neglects the discrete nature of the
lattice. Thus, it is fully ignored that fracture of materials is ultimately
caused by bond breaking processes on the atomic scale.

A way to understand the processes is to perform
numerical experiments, since experimental information on this length scale is
difficult to obtain. Molecular dynamics
studies have provided useful insight into crack propagation in pure metals and
simple intermetallic alloys, whereas in complex metallic alloys the mechanisms
are not yet so clear. Atomistic studies show for example that cracks remain
stable in a region around the critical stress intensity factor due to the
discrete nature of the lattice. This effect is called lattice trapping
\cite{Thomson71}. A further consequence of the discrete lattice is that the
fracture behaviour in one and the same plane can depend on the crack
propagation {\sl direction} \cite{Gumbsch00}. Such observations cannot be
explained by a simple continuum mechanical description.


\section{MODEL AND METHOD}
\label{modelnmethod}

\subsection{Icosahedral binary model}
\label{ibm}
\begin{figure} [b]
\centering
\includegraphics[width=0.95\linewidth,clip=]{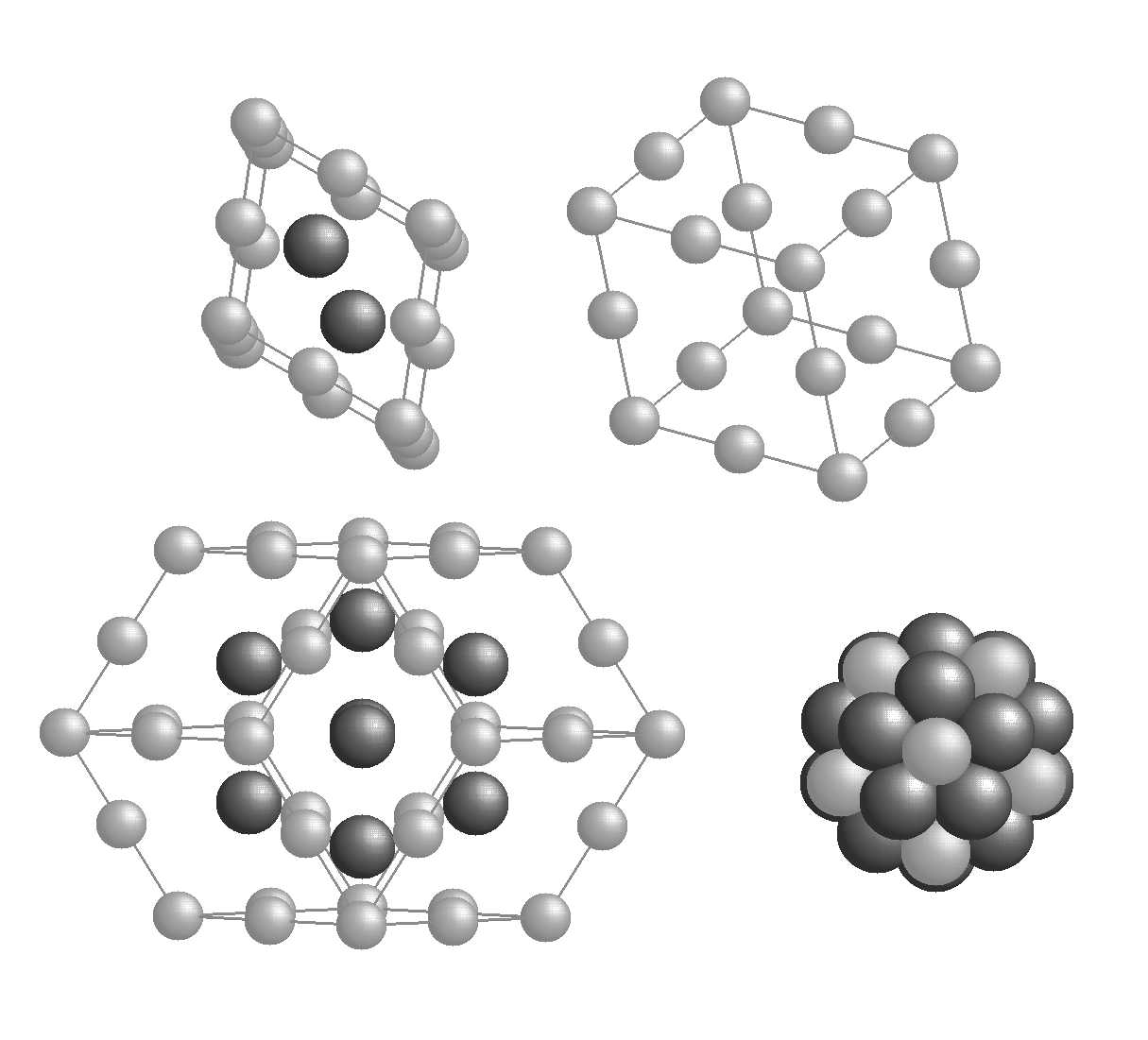}
\caption[ ]{Tiles of the icosahedral binary model decorated with two types of
atoms and the Bergman-type cluster. {\sl Top:} prolate rhombohedron (left) and
oblate rhombohedron (right). {\sl Bottom:} rhombic dodecahedron (left) and
the 45 atoms building the Bergman-type cluster (right) inherent in the model
quasicrystal. Small grey and large black spheres denote A and B atoms
respectively.
\label{fig1.ps}}
\end{figure}

In the numerical experiments we use a three-dimensional model
quasicrystal which has been proposed by Henley and Elser \cite{Henley86} as a
structure model for the icosahedral phase of (Al,Zn)Mg.
This is the simplest possible model quasicrystal that is stabilised by pair
potentials. Furthermore it allows Burgers circuit analysis and is a prototype
of Bergman-type quasicrystals.
As we do not distinguish between Al
and Zn atoms, we term this decoration icosahedral {\em binary} model. It can be
obtained by decorating the structure elements of the three-dimensional Penrose
tiling, the oblate and the prolate rhombohedra (see Fig.~\ref{fig1.ps},
top). Al and Zn atoms (A atoms) are placed on the vertices and the midpoints
of the edges of the rhombohedra. Two Mg atoms (B atoms) divide the long body
diagonal of each prolate rhombohedron in ratios $\tau$:1:$\tau$, where $\tau$
is the golden mean. Two prolate and two oblate rhombohedra with a common
vertex form a rhombic dodecahedron\cite{formingRDs,Roth00}.
To obtain the icosahedral binary model, in these dodecahedra
the atom at the common vertex is removed and the four neighbouring
A atoms are transformed into B atoms. Finally these atoms are shifted to the
common vertex to divide the edges of the corresponding rhombohedra in a ratio
of 1:$\tau$. Fig.~\ref{fig1.ps} (bottom left) shows the final decoration of
these dodecahedra, in which the B atoms form hexagonal bipyramides.
This modification increases
the number of Bergman-type clusters (see Fig.~\ref{fig1.ps}, bottom right)
inherent in the structure, leads to a higher stability with the potentials
used, and takes better into account the experimentally observed stoichiometry
of the quasicrystal. The Bergman-type clusters may also be interpreted as
building units of the quasicrystal and are the
main feature of the structure apart from the plane structure.

\subsection{Molecular dynamics technique}
\label{md}

As is often done in fracture simulations (see e.g.\ Abraham~\cite{Abraham03}) we use simple Lennard-Jones
potentials to model the interactions. The following facts led to this choice:

First, the very few potentials available for quasicrystals are
unsuitable for {\em fracture} simulations: These pair potentials stabilise the
quasicrystal only for a fixed volume. With free surfaces atoms sometimes
simply evaporate (e.g.\ with potentials based on those of Hafner et al.~\cite{Hafner90}).
Very long-ranged potentials with Friedel oscillations (e.g.\ those of
Al-Lehyani et al.~\cite{Allehyani01}) frequently
display such large cohesive energies that nearly no
elastic deformation is possible and instead intrinsic cracks develop.

Second, many simulations have proven that model potentials are helping to
understand the elementary processes in fracture (see e.g.\ Abraham~\cite{Abraham03}) and so are reasonable when
qualitative mechanisms are the centre of interest. For quantitative results, a
known limitation is the neglect of non-local and many-body interactions.

Third, the Lennard-Jones potentials used \cite{Roesch03,Roesch04} keep the
model stable even under strongest mechanical deformations or irradiation
(introduction of point defects) and have  been used in our group in many
simulations of dislocation motion \cite{Schaaf03} or even shock waves\cite{Roth05}. The
structure is robust under a wide variation of the potential depths. Very
similar potentials have shown to stabilise the icosahedral atomic
structure in a simpler model \cite{Roth95}. It is also
known since the early fifties that Lennard-Jones potentials favour icosahedral
clusters \cite{Frank52}, indicating that these potentials are useful for
structures like icosahedral quasicrystals.

The minima of the potentials for interactions between atoms of the same type
are set to $\epsilon_{0}$, whereas the minimum of the potential for the
interactions between atoms of different kind is set to $2\epsilon_{0}$. 
The conclusions drawn from our simulations, however, remained essentially
unchanged if all binding energies are set equal. The shortest distance between
two A atoms is denoted $r_{0}$. All masses are set to $m_{0}$. The time is
then measured in units of $t_{0}=r_{0}\sqrt{m_{0}/\epsilon_{0}}$.

The molecular dynamics simulations were carried out using the microcanonical
ensemble with the program code IMD \cite{Stadler97,imd}. It performs well on a large
variety of hardware, including single and dual processor workstations and
massively parallel supercomputers.

First, we searched for the potential cleavage planes. According to the Griffith
criterion they should be planes of low surface energy
\cite{griffithcut}. To identify these surfaces we
relax a specimen and split it into two parts. Subsequently, the two regions are
shifted apart rigidly. The surface energy is then calculated from the
energy difference of the artificially cleaved and the undisturbed
specimen.

\begin{figure} [t]
\centering
\includegraphics[height=0.95\linewidth,angle=-90,clip=]{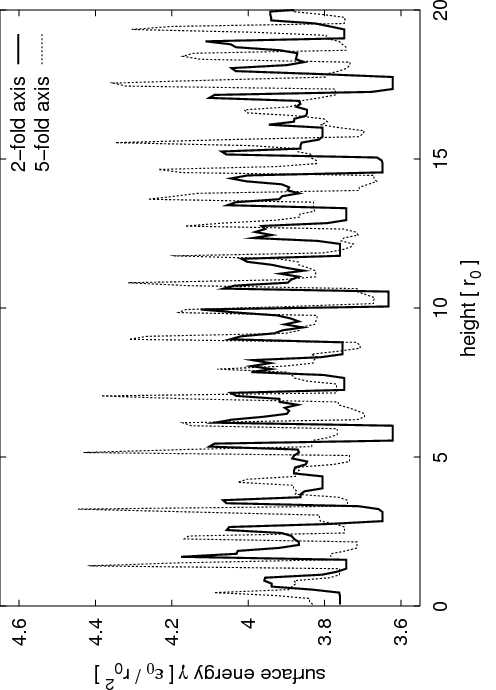}
\caption[ ]{Surface energy of cleavage planes perpendicular to two- and
  fivefold axes.
\label{fig2.ps}}
\end{figure}

In contrast to simple periodic crystals, the atomistic structure of the planes
and therefore also the surface energy in quasicrystals varies strongly within
the family of planes perpendicular to a fixed axis.
In the present model, surfaces with low interface energy are located
perpendicular to two- and fivefold directions at certain heights
(Fig.~\ref{fig2.ps}).  Perpendicular to other
directions the plane structure is less pronounced and the minimal surface
energies are higher.

\begin{figure} [b]
\centering
\includegraphics[width=0.95\linewidth,clip=]{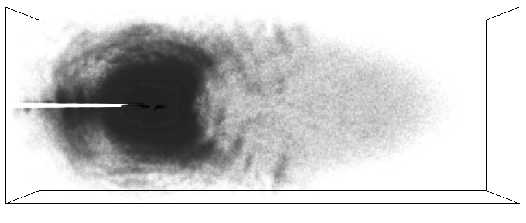}
\caption[ ]{Kinetic energy density. Sound waves emitted by the propagating 
  crack and the elliptical region without damping are clearly visible. See
  online movie~\cite{moviesw}.
\label{fig3.ps}}
\end{figure}

Since we are interested in
the morphology of fracture surfaces we apply a sample form that allows us to follow
the dynamics of the running crack for a long time. For this purpose, a strip
geometry is used to model crack propagation with constant energy release rate
\cite{Gumbsch97}. The samples consist of about 4 to 5 million atoms, with
dimensions of approximately $450r_{0} \times 150r_{0} \times 70 r_{0}$. 
Periodic boundary conditions \cite{approx} are applied in the direction
parallel to the crack front. For the other directions, all atoms in the
outermost boundary layers of width $2.5 r_{0}$ are held fixed. An atomically
sharp seed crack is inserted 
at a plane of lowest surface energy, from one side to about one
quarter of the strip length. The system is uniaxially strained perpendicular
to the crack plane up to the Griffith load and is relaxed to obtain the
displacement field of a stable crack at zero temperature.
Then a temperature of about $10^{-4}$ of the melting
temperature is applied \cite{tempcontrol} to the configurations with and
without the relaxed crack.
From the resulting configurations we obtain an averaged
displacement field for this temperature.
The crack now is
further loaded by linear scaling of this displacement field. The response of
the system then is monitored by molecular dynamics techniques. The sound
waves emitted by the propagating crack (see Fig.~\ref{fig3.ps} and online
movie~\cite{moviesw}) are damped away outside of an elliptical stadium
\cite{Gumbsch97} to prevent reflections.

\subsection{Visualisation}
\label{vis}
To study crack propagation on an atomic level the selection and reduction of
data is of crucial importance. Due to the large number of atoms required 
for the study of crack propagation in three-dimensional systems, it is not
feasible to
write out the positions of all atoms during the simulation, and even less to
display all of them. To obtain a first insight into crack propagation only atoms
near the fracture surfaces are of interest. 
Whereas they can be visualised in periodic crystals by plotting only
those atoms whose potential energy exceeds a certain threshold, this technique
is not applicable for quasicrystals.
Because of the largely varying environments the
potential energy varies significantly from atom to atom, even for atoms
of the same type in a defect-free sample. 

\begin{figure} [t]
\centering
\includegraphics[width=0.95\linewidth,clip=]{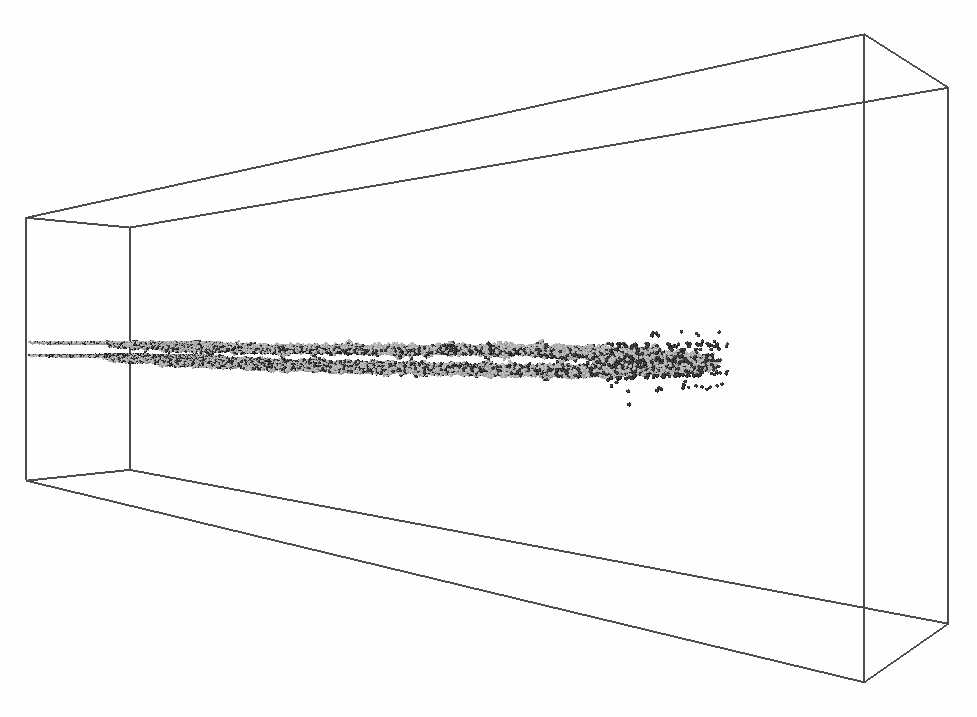}
\caption[ ]{Snapshot of a simulation with some 4 million atoms. Only atoms
  with low coordination number are displayed. See online movie~\cite{moviedf}.
\label{fig4.ps}}
\end{figure}

A solution to this problem is to display only those atoms whose coordination 
number is smaller than a certain threshold. This number is
evaluated by counting atoms within the nearest neighbour distance. Like
the potential energy, in quasicrystals the coordination number varies from
atom to atom, but to a much smaller degree. As atoms near defects have a
significantly lower coordination number, it becomes possible to visualise
fracture surfaces and dislocation cores. For the A atoms the threshold for the
coordination number is set to 12, whereas for the B atoms it is set to
14. With this method, the number of atoms to write out or to display can be
reduced by three orders of magnitude.
Fig.~\ref{fig4.ps} shows a snapshot
of a simulation with some 4 million atoms filtered by this technique, which
was also used in a movie that is available online~\cite{moviedf}.

As can be seen from Fig.~\ref{fig4.ps} the fracture surfaces are rough.
Therefore to decide to which fracture surface an atom belongs we apply the
displacement vectors between the initial configuration with the built-in seed
crack and the fractured sample.
The morphology of the fracture planes is then analysed via geometrical
scanning of the atoms forming the surfaces. The roughness can be visualised by
colour coding the height of the surface in a view normal to the fracture
surface.

\begin{figure} [t]
\centering
\includegraphics[width=0.95\linewidth,clip=]{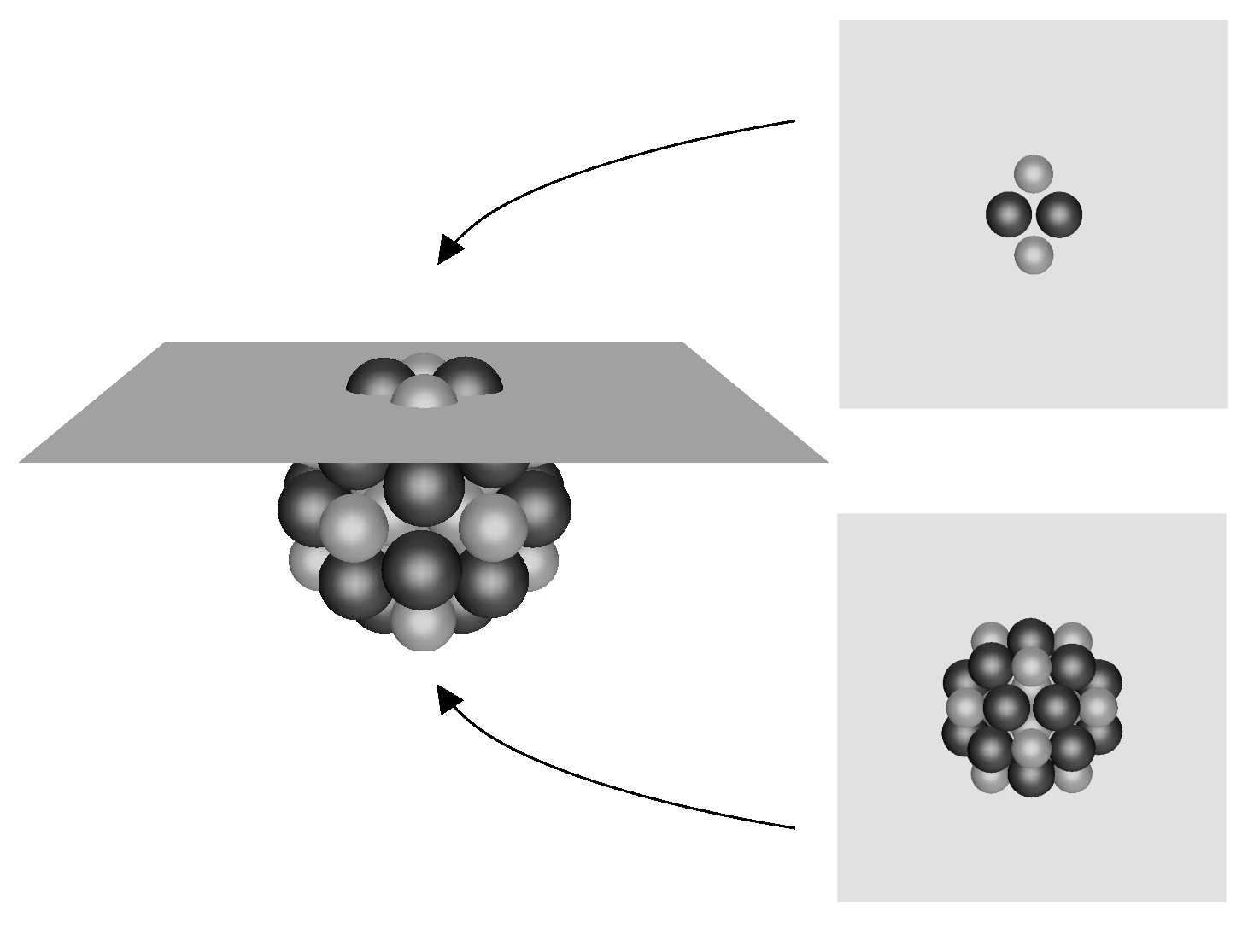}
\caption[ ]{Bergman-type cluster cut by a flat surface.
\label{fig5.ps}}
\end{figure}

To investigate the influence of the Bergman-type clusters on cleavage they
have to be displayed together with the fracture surfaces intersecting
them. This is done by restoring the initial sample without crack at zero
temperature. The atoms forming the two sample parts are taken back to their
positions in this initial sample and then scanned geometrically. In addition
all atoms forming clusters in the vicinity of this surfaces are known. By
displaying only these atoms and the scanned surface one directly can see where
and to which amount clusters are cut. A problem of this kind of visualisation
is shown in Fig.~\ref{fig5.ps}, where a cluster is cut by a flat
surface. When looked-at from above it is obvious that only four atoms are
separated from the rest of the cluster. When looked-at from below one could
get the impression that the cluster is
heavily intersected. On a real fracture surface clusters with centres above
and below the crack surface are present. Therefore both views of
Fig.~\ref{fig5.ps} are appearing at the same time in a two-dimensional
projection. Thus such pictures are not very intuitive.

\begin{figure} [b]
\centering
\includegraphics[width=0.95\linewidth,clip=]{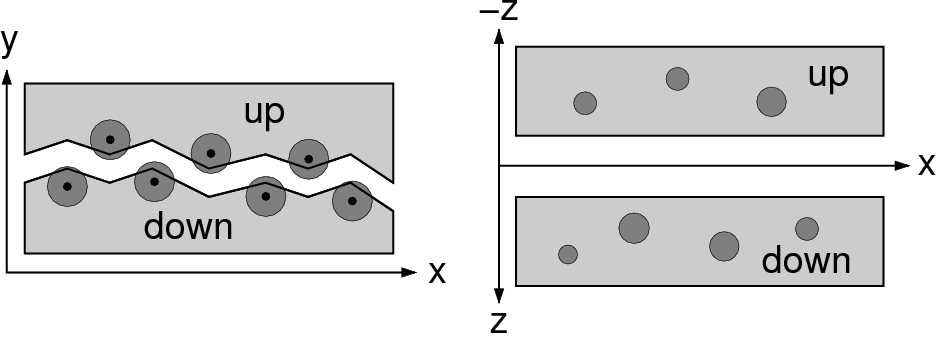}
\caption[ ]{Visualisation of the clusters cut by the dynamic crack. Midpoints
  of clusters are indicated as black dots, the clusters are idealised as
  spheres. The crack propagated from the left to the right. The upper and
  lower fracture surfaces are projected onto an $x-z$ plane.
\label{fig6.ps}}
\end{figure}

A way out of this dilemma is presented in Fig.~\ref{fig6.ps}. Clusters
with midpoints above the crack surface are displayed together with the upper
geometrically scanned fracture surface only, the other clusters are shown together
with the lower fracture surface. As a result we get two pictures with clusters
cut by surfaces. Note that for a qualitative and quantitative analysis always
both pictures or sets of data are needed.

\section{RESULTS}
\label{results}
In this section the results of the numerical simulations are presented. For
practical purposes we define $k$ as the stress intensity factor $K$ relative to
the stress intensity factor $K_{G}$ at the Griffith load:
$$k=K/K_{G}.$$
The orientations of the samples are characterised by the notation $y_{x},$
where $y$ is an axis perpendicular to the cleavage plane and $x$ is an axis in
the crack propagation direction (see Fig.~\ref{fig6.ps}). An axis
perpendicular to a fivefold (5) and a twofold (2) axis is denoted
pseudo-twofold (p2) axis.

\subsection{Crack velocities}
\label{vel}
\begin{figure} [t]
\centering
\includegraphics[height=0.95\linewidth,angle=-90,clip=]{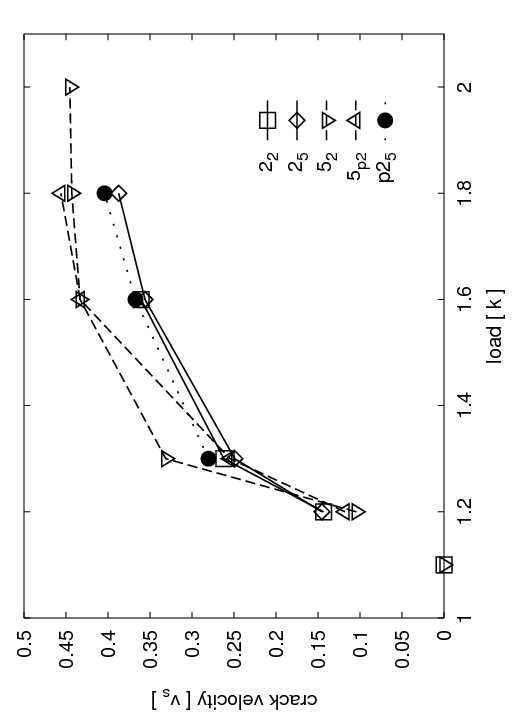}
\caption[ ]{Average crack velocities for different loads and orientations.
\label{fig7.ps}}
\end{figure}

Simulations were performed for different orientations with loads in a range
from $k=1.1$ to $k=2$ (see notations and loads in
Fig.~\ref{fig7.ps}). Brittle fracture without any crack tip plasticity is
observed irrespective of the orientation of the fracture plane. For loads
below $k=1.2$, the crack propagates only a few atomic distances, and then
stops. Hence the energy needed for initiating crack propagation is about 1.4 times
the value predicted by the Griffith criterion. Therefore, a simple global
thermodynamic description of fracture is not applicable.
The minimal velocity for brittle crack propagation is about 10\% of the
shear wave velocity \cite{isotropic} $v_{s}$. For loads $k \ge 1.2$ the
velocity increases monotonically with the applied load. The crack velocities
are in a range of 10-45\% of $v_{s}$ (see Fig.~\ref{fig7.ps}).
At high loads the crack velocities on fivefold cleavage planes show higher average
velocities than on the other planes.
Velocities for the two different crack propagation directions on the fivefold
planes differ significantly at intermediate loads ($k = 1.3$).
This coincides with ledges that are produced in the fracture surface (see
section \ref{surf} and Fig.~\ref{ORIENTATIONS.ps}, bottom).

\subsection{Fracture surfaces}
\label{surf}
\begin{figure*} [p]
\centering
\includegraphics[width=0.85\linewidth,clip=]{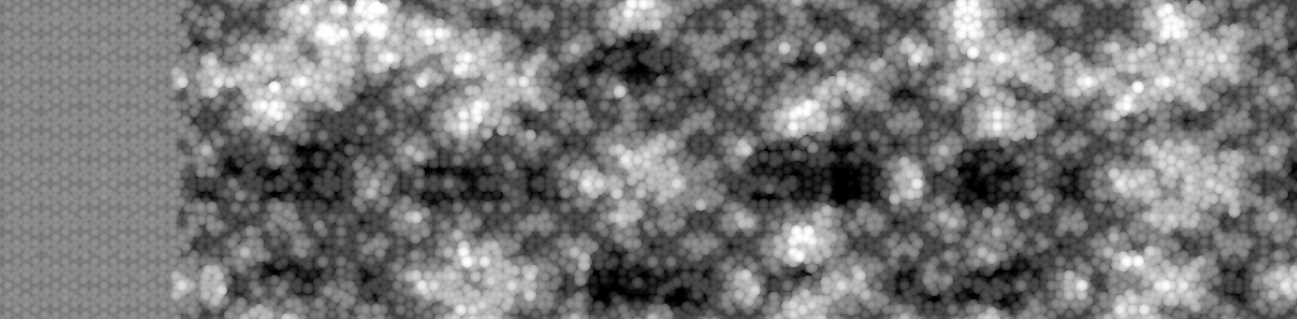}
\includegraphics[width=0.85\linewidth,clip=]{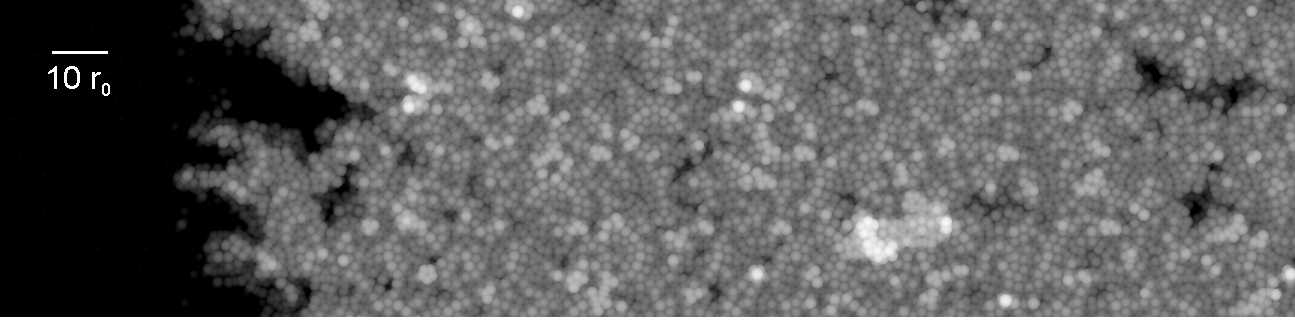}
\includegraphics[width=0.85\linewidth,clip=]{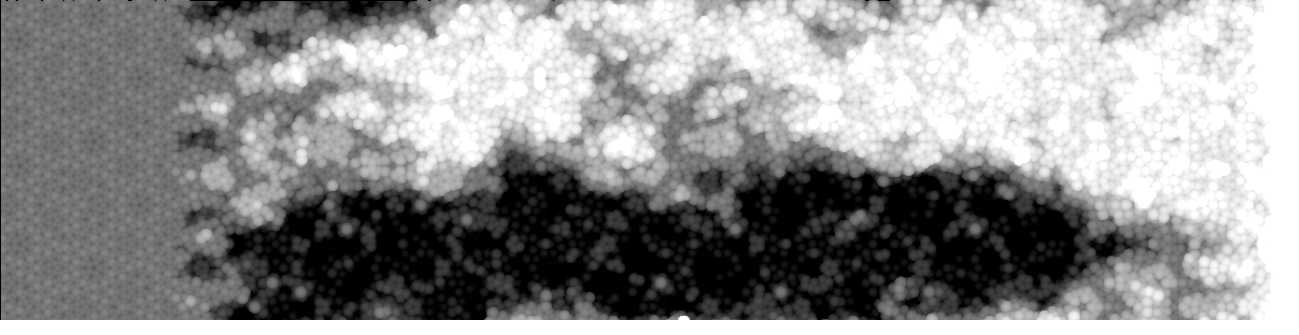}
\caption[ ]{Height profiles of sections of the simulated fracture
  surfaces. Load: $k=1.3$, orientation $2_{2}$ (top), $5_{2}$ (middle),
  $5_{\text{p}2}$ (bottom). The height increases from black ($\le -2r_{0}$) to
  white ($\ge +2r_{0}$). The scanning sphere has the same size as an atom of
  type B.
\label{ORIENTATIONS.ps}}
\end{figure*}

To analyse the morphology of the fracture surfaces, they are
displayed as described in section~\ref{vis}. In Fig.~\ref{ORIENTATIONS.ps} 
the average height is shown in grey, heights above $+2r_{0}$ are shown
in white and heights below $-2r_{0}$ are shown in black. The crack propagation
direction is from the left to the right. The initial fracture surface is flat,
as can be seen from the homogeneous regions on the left. The surfaces
resulting from the propagation of the crack, however, show pronounced patterns
of regions with different heights. From the observation of the fracture
surfaces it is already evident that they are rough and that the
peak-to-valley roughness is of the order of the diameter of the clusters. The
peak-to-valley roughness and the root-mean-square roughness of the height
profiles both increase~\cite{rough} for higher loads for surfaces without ledge
formation. For fracture surfaces perpendicular to twofold and fivefold axes
the crack fluctuates about a constant height (in the areas without ledges). In
contrast, a crack inserted perpendicular to a pseudo-twofold direction seems to
deviate from this plane \cite{deviation}.

\begin{figure*} [p]
\centering
\includegraphics[width=0.85\linewidth,clip=]{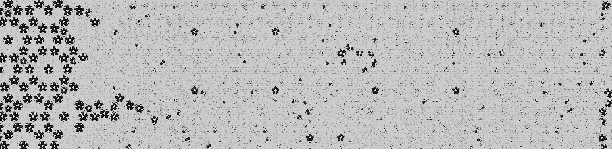}
\includegraphics[width=0.85\linewidth,clip=]{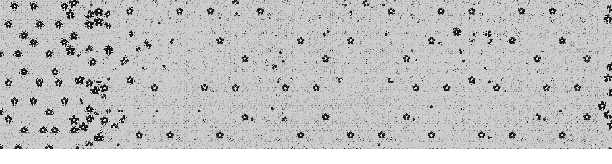}
\caption[ ]{Clusters cut by the fracture surfaces. The visualisation
  technique is described in Sec.~\ref{vis} and
  Fig.~\ref{fig6.ps}. Load: $k=1.3$, orientation: $5_{2}$.
\label{C52.ps}}
\end{figure*}

\subsection{Anisotropy}
\label{anisotropy}
As can be seen already from the fracture surfaces perpendicular to a fivefold
axis in Fig.~\ref{ORIENTATIONS.ps} crack surfaces for the same cleavage plane
differ significantly for different in-plane crack propagation directions. For
the orientation $5_{\text{p}2}$ ledges are produced, while no ledges form for
the orientation $5_{2}$. By visual inspection of the fracture surfaces (see
Fig.~\ref{ORIENTATIONS.ps}) and evaluation of height-height-correlation
functions \cite{Roesch03} it becomes evident that for every orientation
there are distinct angles in the height profile, which show pronounced height
correlations that correspond to markings in the fracture surfaces or to
ledges. These angles are given in Table~\ref{angle}.

\begin{table} [b]
\caption[]{\label{angle}
Angles observed in the height profiles of the fracture surfaces. Angles
measured clockwise from the crack propagation direction get a negative sign.}
\begin{ruledtabular}
\begin{tabular}{lccccc}
orientation & $2_{2}$ & $2_{5}$ & $5_{2}$ & $5_{\text{p}2}$ & $\text{p}2_{5}$\\
angles&
$0^{\circ}, \pm 32^{\circ}$&
$0^{\circ}, + 32^{\circ}, - 58^{\circ}$&
$0^{\circ}, \pm 36^{\circ}$&
$\pm 18^{\circ}$&
$0^{\circ}, \pm 90^{\circ}$\\
\end{tabular}
\end{ruledtabular}
\end{table}

\subsection{Clusters}
\label{clusters}

In Fig.~\ref{C52.ps} the clusters cut by the fracture surfaces are presented
as described in section \ref{vis} and Fig.~\ref{fig6.ps}. It is obvious
from Fig.~\ref{C52.ps} that the dynamic crack does not perfectly circumvent the
clusters, but intersects them partially (right side of Fig.~\ref{C52.ps}).
These intersections, however, are much less frequent than for the flat seed
cracks (left side of Fig.~\ref{C52.ps}). More detailed analyses for different
orientations validate this statement. For the orientations perpendicular to
twofold and fivefold axes at $k=1.3$ the ratio of clusters cut by the dynamic
crack to clusters intersected by flat cuts is approximately 0.6. Additionally
the absolute value of clusters cut by the crack for the fivefold surfaces is
lower than for the twofold surfaces.

\begin{figure} []
\centering
\includegraphics[width=0.95\linewidth,clip=]{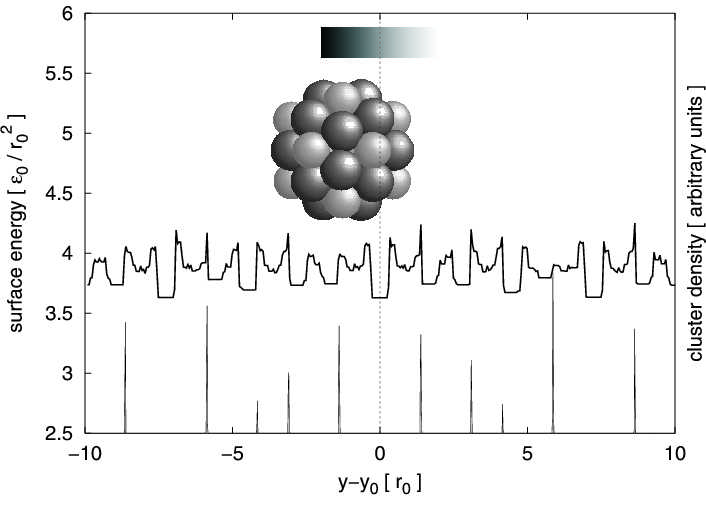}
\caption[ ]{Surface energy and density of cluster centres for the orientation
  $2_{2}$. The corresponding fracture surface is shown in
  Fig.~\ref{ORIENTATIONS.ps}, top.
\label{fig10.ps}}
\end{figure}

\begin{figure} []
\centering
\includegraphics[width=0.95\linewidth,clip=]{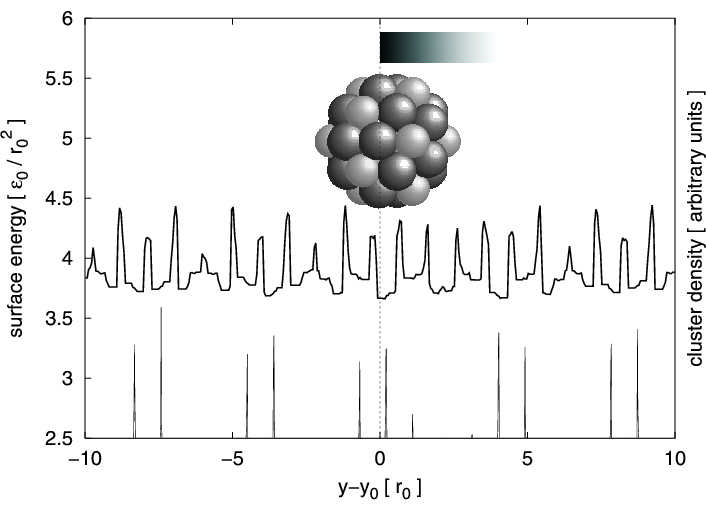}
\caption[ ]{Surface energy and density of cluster centres for the orientation
  $5_{2}$. The corresponding fracture surface is shown in
  Fig.~\ref{ORIENTATIONS.ps}, middle.
\label{fig11.ps}}
\end{figure}

Fig.~\ref{fig10.ps} and Fig.~\ref{fig11.ps} display bottom up:
The density of the cluster centres, the surface energy, a cluster in the
proper length scale, the grey coding of the heights in
Fig.~\ref{ORIENTATIONS.ps} (top, middle), and the position of the seed crack
(dashed vertical line). For the
twofold fracture surface the low energy seed crack is located between two
peaks in the cluster density, whereas for the fivefold surface this seed crack
is situated close to a peak of this density. It is evident from the figures
that it is not possible to circumvent all clusters by a planar cut. 


The grey coding is adjusted to the average height of the fracture surfaces. It
is therefore evident from Fig.~\ref{fig11.ps} that the crack deviates for the
orientation $5_{2}$ from the low energy cleavage plane of the seed crack to a
parallel plane, reducing the number of cluster intersections dramatically (see
also Fig.~\ref{C52.ps} and Fig.~\ref{ORIENTATIONS.ps}, middle).
Samples cut flat at the new height show slightly higher surface energy (see
also Fig.~\ref{fig11.ps}).
However, for low loads and low roughness the actual fracture
surfaces of the dynamic cracks have about 5-15\% higher energies than those of
the low energy seed cracks. To assure that the dynamic crack is departing from
the initial plane not in a random manner the seed crack was built in at the
position colour coded as medium grey in Fig.~\ref{fig11.ps}. The
resulting fracture surface had a similar roughness but the crack did not
change to a parallel plane.

\section{DISCUSSION}
\label{discussion}
Taken together the results of our simulations presented above indicate that
the distribution of the clusters is crucial for the fracture behaviour:
First, circumventing the clusters or intersecting them disturbs the
propagating crack and leads to additional radiation. This manifests itself in
the crack speed. Dynamic cracks propagating in fivefold planes with few
cluster intersections are faster than those in
twofold planes, where the absolute number of cluster intersections is
higher. Also the generation of ledges for low loads in the orientation
$5_{\text{p}2}$ costs energy and therefore slows down the crack even further.
Second, circumvention of the clusters leads to characteristic
height-variations. Therefore the peak-to-valley roughness of the fracture
surfaces is of the order of the diameter of the clusters.
Third, the observed patterns in the fracture surfaces correspond to lines
along which the clusters are located. The associated angles are given by the
icosahedral symmetry of the sample, namely
$0^{\circ},18^{\circ},31.72^{\circ},36^{\circ},58.28^{\circ},$ and
$90^{\circ}$ (see Table~\ref{angle}).
Ledges seem to be produced only for the smallest angles measured from the
crack propagation direction.
Fourth, less clusters are intersected by the fracture surfaces than by the
flat seed cracks.
Fifth, a seed crack at a low energy cleavage plane deviates to a parallel
plane to reduce the number of cluster intersections in spite of the higher
energy required to form the fracture surfaces. In contrast, a crack
built-in at this new position does not show such a deviation.

Another observation of the simulations is that the plane structure of the
quasicrystal also influences fracture. The fracture surfaces that
are located perpendicular to the twofold and fivefold symmetry axes show
constant average heights.

The three-dimensional quasicrystals give perfect cleavage fracture with no
indication of any dislocation activity. This is in contrast to results on
two-dimensional decagonal quasicrystals, where a dislocation enhanced fracture
mechanism has been observed \cite{Mikulla98}. However a corresponding
three-dimensional decagonal quasicrystal would have a periodic direction with
a straight dislocation line. In the simulations presented here this direction
is also quasiperiodic. As the clusters have a strong influence on fracture
they also may bend and hinder dislocation lines. So dislocation
emission in the three-dimensional icosahedral quasicrystal modelled here
should be less likely than in the two-dimensional decagonal model
quasicrystal. Very high stresses are indeed needed to move
dislocations in our model quasicrystal in molecular dynamics simulations
\cite{Schaaf03}.

There are also indications for the stability of the clusters from the
electronic structure of quasicrystals:
First, experiments and ab-initio calculations show that directional
bonding may be present within clusters of quasicrystals
\cite{clustersinbulk,Kirihara03}. Second, the electrons may additionally
stabilise the
clusters because of their hierarchical structure \cite{Janot94}. Therefore
they should be even more stable than we have modelled them with
simple pair potentials. With this evidence the results concerning the clusters
seem reasonable and should even underestimate their stability. 

\begin{figure} []
\centering
\includegraphics[width=0.95\linewidth,clip=]{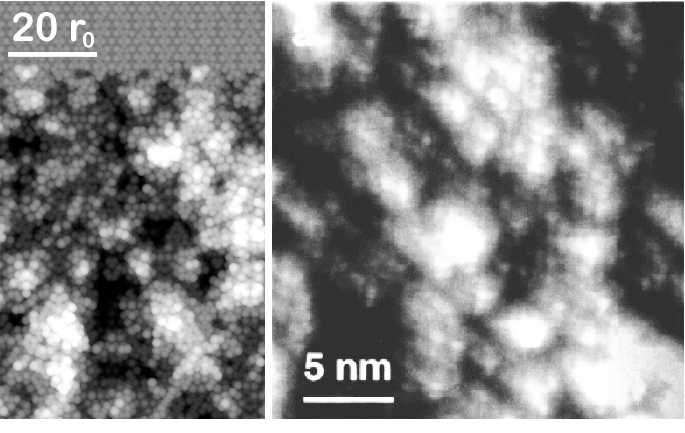}
\caption[ ]{Fracture surfaces perpendicular to twofold axes. In the left
  picture (simulation) the atomically sharp seed crack can be seen on the top,
  whereas below this area the simulated fracture surface appears. The
  orientation of the sample is $2_{2}$, the load was $k=1.3$. The surface was
  geometrically scanned with an atom of type
  B. Thus atomic resolution is achieved. The right picture (experiment,
  adopted from Ebert et al.~\cite{Ebert98}) shows an STM-image of
  icosahedral Al-Pd-Mn cleaved in ultrahigh vacuum. As $20r_{0}$ is
  approximately 5~nm the surfaces are displayed at the same length scale.
\label{fig12.ps}}
\end{figure}

So far fracture experiments in ultrahigh vacuum have only been
performed on icosahedral Al-Mn-Pd, which has a more complicated atomic
structure than the icosahedral binary model. Additionally the clusters are not
Bergman-type. Therefore we cannot expect to represent this structure on an
atomic level, when comparing experiments in Al-Mn-Pd with our simulations.
Nevertheless, the size of the clusters, the icosahedral symmetry, and
a distinct plane structure are common features and qualitative aspects should be
reproduced well, namely the size and shape of the patterns and the appearance
of distinct angles on the fracture surfaces. This is indeed the case, as can
be seen in Fig.~\ref{fig12.ps}. There a geometrically scanned fracture
surface generated in our simulations is confronted to an STM-image of Ebert et
al.~\cite{Ebert96,Ebert98} at the same length scale.
As we were able to correlate the observed structures to the clusters in our
model, the similarities corroborate the assumption that the clusters
are responsible for the globular structures observed in experiment.
More detailed comparisons to fracture experiments in icosahedral
Al-Zn-Mg-type quasicrystals would be desirable, but such comparative data is
currently not available.

We also performed fracture simulations in a C15 Laves phase with the same
Lennard-Jones potentials to give further evidence, that the roughness is
correlated to the clusters. This C15 structure is built up from deformed
prolate rhombohedra, one of the major tiles of the quasicrystal (see
Fig.~\ref{fig1.ps} and Sec.~\ref{ibm}). However, no Bergman-type clusters are
present. When colour coded like the quasicrystal, fracture surfaces of the C15
cracks lack any roughness at low loads and cleave smoothly on high symmetry
planes \cite{Roeschtbp}.

When a crack traverses a solid, it leaves a typical non-equlibrium
surface. Thus, up to this point, we have not dealt with any equilibrium
surface. Even the flat Griffith cuts we introduced to perform the numerical
experiments are no explicit equilibrium surfaces~\cite{griffithcut} by
definition. It is known that i-Al-Pd-Mn surfaces sputtered and annealed up to
about 900~K are rough with cluster-like protrusions. Fivefold surfaces annealed
at higher temperatures exhibit flat
terraces~\cite{Schaub94,Gierer97,Cappello02}, which are believed to be bulk
terminated~\cite{Papadopolos02}.
As fivefold surfaces are not as rough as twofold surfaces, they are
often studied in experiment.
Adsorption on these surfaces, nevertheless, is often very
site-sensitive. A recent review on quasicrystal surfaces was given by McGrath
et al.~\cite{McGrath02}.
All of these observations are consistent with our simulations of
non-equilibrium surfaces. Fivefold surfaces experience less roughness than
twofold surfaces (also in agreement with the experiments of Ebert et
al.~\cite{Ebert96,Ebert98}; see Fig.~\ref{ORIENTATIONS.ps}), where more
clusters are intersected. Additionally, since clusters are cut in the
simulations, their binding energy is not so large as to term them
supermolecules. Thus annealing at very high temperatures can favour flat
surfaces~\cite{Cappello02,McGrath02}. Nevertheless, selective adsorption may
then be related to completion of clusters.



\section{CONCLUSIONS}
\label{conclusions}
We have simulated crack propagation in an icosahedral model
quasicrystal. Brittle fracture without any signature of dislocation emission
is observed. The fracture surfaces are rough on the scale of the clusters and
show constant average heights for orientations perpendicular to twofold and
fivefold axes. Thus both the plane structure and the clusters
play an important role in fracture. The influence of the clusters is
also seen in the average crack velocities for different orientations, the
observed patterns in the fracture surfaces, the anisotropy with respect to the
in-plane propagation direction, and the smaller amount of clusters
cut by the propagating crack than by planar cuts. The clusters, too, are a reason
why the positions of the cleavage planes cannot be predicted by a simple
energy criterion.
Since partial cluster intersections occur, the binding energy of the clusters
is not so large as to term them supermolecules.
Nevertheless our observations clearly show that they are not only
structural units but physical entities.


\acknowledgements
Financial support from the Deutsche Forschungsgemeinschaft under contract
numbers TR 154/13 and TR 154/20-1 is gratefully acknowledged.




\begin{thebibliography}{99}
\bibitem{formingRDs}
In rings of oblate rhombohedra the number of clusters is maximised. No
overlapping rhombic dodecahedra are generated. Remaining oblate rhombohedra
stay unchanged.
\bibitem{griffithcut}
To simulate fracture with molecular dynamics it is necessary to determine the
optimal place for a low energy seed crack and to find out the critical strain
for setting the crack into motion. This is always done by determining the
energies of flat cuts. Note, that these do not necessarily represent
equilibrium surfaces which may be bent and curved.
\bibitem{approx}
As quasicrystals show non-periodic translational order we generate
periodic approximants to apply periodic boundary conditions. The 4 to 5
million atoms of our sample then form the unit cell. Because of this high
number of atoms the configuration should mechanically behave like the perfect
quasicrystal.
\bibitem{moviesw}
Cracks propagate by breaking bonds between atoms. In this process, energy is
dissipated in the form of acoustic waves, as clearly visible in the
online movie. The kinetic energy density is colour coded. Regions of
high kinetic energy density are shown in red whereas blue indicates areas of
lower kinetic energy density.
\bibitem{moviedf}
Movie (available online) of a cracking model quasicrystal at a high load
($5_{p2}$, $k=1.6$). By displaying only atoms with a low coordination number
fracture surfaces are visualised. Blue (black in Fig.~\ref{fig1.ps}) and red
(grey in Fig.~\ref{fig1.ps}) balls correspond to the atoms in the icosahedral
binary model. Atoms on the border, which are not allowed to move, are
coloured in yellow and green. On the left the atomically sharp seed
crack can be seen. Due to the high deformation near the crack tip, blue atoms
loose their neighbours and show up for short instances.
\bibitem{rough}
The height profiles were determined via geometrical scanning of the fracture
surfaces with an atom of type B at equidistant
points separated by \begin{math} 0.2 r_{0} \end{math}. For these meshes and
e.g.\ for the orientation \begin{math} 5_{2} \end{math} the root-mean-square
roughness increases from \begin{math} 0.39 r_{0} \end{math} at a load of
\begin{math} k = 1.3 \end{math} (see Fig.~\ref{ORIENTATIONS.ps}, middle) to
\begin{math} 0.54 r_{0} \end{math} for  \begin{math} k = 1.6 \end{math}.
The peak-to-valley roughness increases likewise from 
\begin{math} 3.7 r_{0} \end{math} to \begin{math} 5.1 r_{0}
\end{math}. Although the exact values may depend on the stepwidth (resolution),
scanning sphere size, and scanned area a general tendency for an increased
roughness for increased loads is evident.
\bibitem{deviation}
The question whether this crack finally chooses a plane perpendicular to a
twofold or fivefold axis cannot be answered, as simulations with the
required sample-size would exceed present computer capabilities.
\bibitem{clustersinbulk}
Only clusters in the bulk give reliable information on their relative
stability. In ab-initio studies, where the number of atoms is limited, one
therefore is restricted to small approximants.
\end{thebibliography}
\end{document}